\documentstyle[multicol,aps,prl,epsf]{revtex}
\begin{document}
\draft


\title{Incipient failure in sandpile models}

\author{Onuttom Narayan}
\address{Department of Physics, University of California, Santa Cruz,
CA 95064}
\author{Sidney R. Nagel}
\address{The James Franck Institute and Department of Physics,
University of Chicago, Chicago, IL 60637}

\date{\today}

\maketitle

\begin{abstract}
Elastoplastic and constitutive equation theories are two approaches based
on very different assumptions for creating a continuum theory for the
stress distributions in a static sandpile.  Both models produce the same
surprising prediction that in a two dimensional granular pile constructed
at its angle of repose, the outside wedge will be on the verge of failure.
We show how these predictions can be tested experimentally.
\end{abstract}
\begin{multicols}{2}
\narrowtext
\noindent {\bf 1.Introduction}

Understanding the stress distributions in static piles is a basic and
open question in the field of granular materials. If one tries to
construct a continuum coarse-grained description of such systems, the
difficulty is that, unlike for elastic materials, there is no obvious
relation connecting the stresses to any strain fields. In two
dimensional systems, there are three variables, the components of the
symmetric stress tensor, but only two force balance equations in the
two orthogonal directions. If the three components of the stress
tensor can no longer be expressed in terms of the two components of a
displacement field, one needs one additional equation to obtain the
stresses.  In general three dimensional systems, the situation is even
worse: one has six components of the stress tensor instead of three
components of a displacement field, so that there are three `missing'
equations.

One approach for understanding the general structure of the stress
distributions in granular systems is in terms of a grain by grain
statistical method~\cite{liu}.  However, one would hope that, on length
scales much larger than the individual
grain size, it should be possible to construct a continuum theory
for a static pile of granular matter (a ``sandpile").  Several very
different approaches for constructing such a theory have been tried.  Two
particularly interesting recent models are the elastoplastic theories
epitomized in the work of Cantelaube and Goddard~\cite{goddard} and the
constitutive equation models introduced by Bouchaud {\it et
al.\/}~\cite{bcc}, and subsequently
generalized by Wittmer {\it et al.\/}~\cite{cates}.   Despite their very
different assumptions, both approaches obtain very similar results for
the stress distribution in a two-dimensional triangular sandpile
obtained by pouring grains from a point source. In particular, they
predict that for such a sandpile
there is a wide wedge of material near the surface of the pile that is
on the threshold of failure, in that it is unstable to the application
of further shear stress.

On the face of it, the result of a large wedge at the instability
threshold predicted by both approaches is surprising since there is a
great deal of experience with the flow caused by tilting a pile at its
angle of repose.  In such avalanches the motion is confined only to
grains in a thin boundary layer near the
surface~\cite{Jaeger}~\cite{caveat1}. From such observations one might
have naively expected that only the surface of the sandpile, and not an
entire wedge, would be unstable.

These are dramatic predictions that are inescapable consequences of the
assumptions of these two types of theories.  It is important to see if
experiments can be designed to detect the unstable regions. The purpose
of the present paper is to explore the general consequences of having a
wedge of plastic material near the surface of the pile, and show what
experiments can be performed to probe its existence.

In the constitutive equation models~\cite{bcc}~\cite{cates} it is suggested
that
the components of the stress tensor are actually related to one
another. As mentioned above, in two dimensions one needs one such
``constitutive
equation'' interrelating the stresses; in general three-dimensional
systems one needs three constitutive equations, which can be reduced in
systems with special symmetry. Although the constitutive equations in
general depend on how the system is assembled, the hypothesis that
these equations are local implies that they are determined in any part
of the pile when that part is constructed, and are not affected by
subsequent changes in the loading or by rearrangements in other regions
of the pile. Focusing on a two dimensional sandpile that is built by
pouring grains from a point source, through scaling and symmetry
arguments, Wittmer {\it et al.\/}~\cite{cates} are led to consider
constitutive equations that impose a linear relationship among the
stress components. One is ultimately left with a one-parameter family
of models and solutions thereof.

\begin{figure}
\centerline{
\epsfxsize 4in\epsfbox{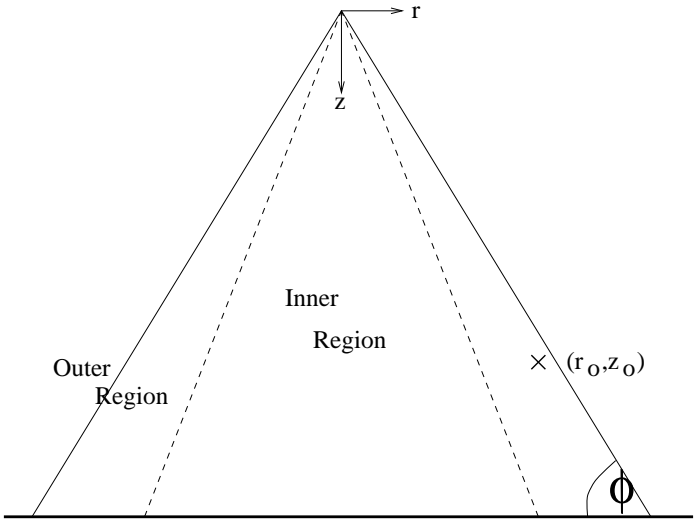}}
\vskip 0.25truecm
\caption{Two dimensional sandpile at its angle of repose. The point
$(r_0,z_0)$ will be used later in the paper.}
\label{fig1}
\end{figure}

For a two dimensional sandpile at its angle of repose, all the models
in the one-parameter family obtain stresses that are linear functions
of the spatial coordinates, and predict that outside a symmetric
triangular ``inner region", the entire pile is on the verge of failure.
This is in the sense that, if $\phi$ is the angle of repose, and $r$
and $z$ are the coordinate directions as shown in Fig.~\ref{fig1}, the
stresses in the outer region satisfy the condition
\begin{equation}
\label{M-C}
{{(\sigma_{rr}-\sigma_{zz})^2+4\sigma_{rz}^2}\over
{(\sigma_{rr}+\sigma_{zz})^2}}=\sin^2\phi
\end{equation}
{\it i.e.\/} the Mohr-Coulomb criterion is saturated. This is
equivalent to the statement that, at {\it any\/} point in the outer
region, one can find a suitably oriented plane for which the ratio of
the shear stress to the normal stress is $\tan\phi,$ which is the
maximum possible value before plastic flow takes place.
\vskip -0.25cm

The stress distribution obtained by Cantelaube and
Goddard~\cite{goddard} is similar to that obtained by Refs.\cite{bcc}
and~\cite{cates}, although these authors start from a completely
different point of view. They assume that a granular material can be
treated as an {\it elastic\/} medium, with the added feature that when
the ratio of the shear strain to the normal strain goes beyond a
critical value, the corresponding ratio of {\it stresses\/} saturates
at $\tan\phi.$ Thus a sandpile can be divided into elastic and plastic
regions, with the stresses expressible in terms of displacement fields
in the elastic regions, and satisfying the Mohr-Coulomb yield criterion
in the plastic regions. For a two-dimensional system, one can eliminate
the strains from the problem and obtain expressions entirely in terms
of stresses. (One still needs to specify the boundaries between the
elastic and plastic regions in order to determine the stresses.)  If
one assumes a scaling form $\sigma_{ij}=z f_{ij}(r/z)$ for the
stresses, one finds that the region adjacent to the sandpile surface
must be plastic. Thus the elastoplastic models~\cite{goddard} and the
ones proposed by  Bouchaud {\it et al.\/}~\cite{bcc} and Wittmer {\it
et al.}~\cite{cates} are in agreement in predicting a wide wedge of
material near the surface of a sandpile at its angle of repose that is
on the threshold of failure.  However a surprising, and we think
counterintuitive, result of the elastoplastic approach~\cite{goddard} is
that even for a triangular sandpile with
arbitrary slope including one {\it well below\/} its angle of repose,
the plastic region must persist if the stresses satisfy the scaling form
above.

In the rest of this paper, we find the destabilizing perturbations that
can experimentally probe the plastic region in these two types of
theories.  Because their approaches are different, they must be
considered separately.  In the scenario of Refs.~\cite{bcc}
and~\cite{cates}, an appropriately chosen infinitesimal perturbation
anywhere in the outer region of the sandpile destabilizes it. We show
this by demonstrating that, when the extra perturbation is included,
there is no solution to the equations for the stresses that satisfies
the Mohr-Coulomb yield criterion everywhere in the pile. The effects of
such a perturbation necessarily extend to the outer surface of the
pile. For elastoplastic theories, an infinitesimal force applied in an
appropriate direction to a section of the bottom of the plastic part of
the pile is sufficient to destabilize it. We discuss how these results
would be modified for a material with a non-zero Bagnold hysteresis
angle.

\noindent{\bf 2. Constitutive equation models}

In the scenario of Refs.~\cite{bcc} and~\cite{cates}, if a localized
body force of magnitude $f$ is applied in the positive $r$ direction at
some point $(r_0,z_0)$ in the outer region (see Fig.~\ref{fig1}), the
equations satisfied by the change $\sigma^\prime$ in the stress tensor
$\sigma$ are
\begin{eqnarray}
\partial_r\sigma^\prime_{rr}+\partial_z\sigma^\prime_{rz}&=&f
\delta(r-r_0) \delta(z-z_0)\nonumber\\
\partial_r\sigma^\prime_{rz}+\partial_z\sigma^\prime_{zz}&=&0
\label{fbal}
\end{eqnarray}
with the boundary conditions that the change in the normal and shear
stress on the sloping sidewalls of the sandpile must be zero. As
discussed in the previous section, the arguments suggesting a local
constitutive equation imply that this equation is not affected by
subsequent changes in the loading, so that the components of the change
in the stress tensor must also satisfy the equation
\begin{equation}
\sigma^\prime_{rr}=\eta\sigma^\prime_{zz}+\mu\sigma^\prime_{rz}
\label{consrel}
\end{equation}
where $\eta$ and $\mu$ are related, so that there is one free parameter
in the equations.~\cite{cates} For $\mu\neq 0,$ Eq.(\ref{consrel}) is
actually two different equations on both sides of the symmetry axis
$r=0,$ leading to a kink in the stress tensor. We shall ignore this at
present, since we shall later see that it is of no consequence.

{}From Eq.(\ref{fbal}b), $\sigma^\prime_{zz}=-\partial_r V$ and
$\sigma^\prime_{rz}=\partial_z V,$ where $V$ is a scalar field.
Substituting in Eq.(\ref{fbal}a), we obtain
\begin{equation}
-\eta\partial_r^2 V +\mu\partial_r\partial_z V +\partial_z^2 V
=\delta(r-r_0)\delta(z-z_0) f.
\label{waveeqn}
\end{equation}
The homogenous version of this equation is an asymmetric wave
equation~\cite{cates} with wave velocities $dr/dz=c_1$ and $-|c_2|.$
Eq.(\ref{waveeqn}) has the solution
\begin{eqnarray}
V(r,z)=\lambda
\Big[\Theta\big(c_1(z&-&z_0)-r+r_0\big)\nonumber\\
   &-&\Theta\big(|c_2|(z_0 - z) -r +r_0\big)\Big]
\label{Vsoln}
\end{eqnarray}
with $\lambda=f/(c_1+|c_2|).$
Therefore we obtain
\begin{eqnarray}
\sigma^\prime_{zz}&=&\lambda
\Big[\delta\big(c_1(z - z_0) - r + r_0\big) - \delta\big(|c_2|( z_0
-z) -r +r_0\big)\big]\nonumber\\
\sigma^\prime_{zr}&=&\lambda\Big[c_1\delta
\big(c_1(z - z_0) -r + r_0\big) +\nonumber\\
&{ }&\qquad\qquad\qquad |c_2|\delta\big( |c_2|(z_0-z) -r +
r_0\big)\Big]\nonumber\\
\sigma^\prime_{rr}&=&\eta\sigma^\prime_{zz}+\mu\sigma^\prime_{rz}.
\label{stress}
\end{eqnarray}

All components of the change in the stress tensor thus consist of a
pulse travelling outward at a velocity $c_1$ and another travelling
inward with a velocity $c_2,$ as one proceeds downward into the pile.
This is as one would expect for a wave equation. The inward pulse is
partly reflected and partly refracted when it reaches the symmetry axis
$r=0,$ emerging as two outward pulses both with speed $c_1.$

Since we wish to demonstrate a violation of the Mohr-Coulomb criterion,
it is sufficient to do this for one of the pulses. Accordingly, we
restrict our attention to the path of the outward propagating pulse.
Along this path, we have
\begin{eqnarray}
\sigma^\prime_{zz}&=&\sigma^\prime_{zr}/c_1\nonumber\\
\sigma^\prime_{rr}&=&c_1\sigma^\prime_{zr}
\label{outstress}
\end{eqnarray}
with $\sigma^\prime_{zr}>0$ for $f>0.$ In the last of these equations,
we have used the relation~\cite{cates} $\eta/c_1+\mu=c_1$ which is valid
for all $\mu.$ When $\sigma^\prime$ is small\cite{caveat3}, the
requirement that the Mohr-Coulomb criterion should be satisfied is
equivalent to the condition
\begin{equation}
{{(\sigma_{zz}-\sigma_{rr})(\sigma^\prime_{zz}-\sigma^\prime_{rr})
+4\sigma_{rz}\sigma^\prime_{rz}}\over{(\sigma_{zz}+\sigma_{rr})
(\sigma^\prime_{zz}+\sigma^\prime_{rr})}}\leq\sin^2\phi.
\label{yield1}
\end{equation}
One can verify that, irrespective of $\mu,$ the unperturbed stresses
satisfy the conditions~\cite{cates} $(\sigma_{zz}-\sigma_{rr})
=2\sigma_{rz}\tan\phi=\sin^2\phi(\sigma_{zz}+\sigma_{rr}).$
Substituting in Eq.(\ref{yield1}), the inequality simplifies to
\begin{equation}
\sigma^\prime_{rz}-(\tan\phi)\sigma^\prime_{rr}\leq 0
\end{equation}
which from Eq.(\ref{outstress}) is equivalent to
\begin{equation}
[1-c_1\tan\phi]\sigma^\prime_{zr}\leq 0.
\end{equation}
Since $\sigma^\prime_{zr}>0$ for $f>0,$ and $c_1\tan\phi <1$ in order
for the existence of an outer region in the sandpile, we see that a
horizontal outward directed force added anywhere in the outer region of
the sandpile results in there being no possible solution to the stress
equations, {\it i.e.\/} the sandpile is destabilized.
\bigskip

\noindent{\bf 3. Elastoplastic models}\\
\indent In the scenario of Ref.~\cite{goddard}, one assumes that the 
sandpile
can be divided into elastic regions, where the stresses are related to
underlying strain fields, and plastic regions where Eq.(\ref{M-C}) is
satisfied. The specific solution for a sandpile at its angle of repose
in Ref.~\cite{goddard} has an outermost plastic region in which the
stresses are linear functions of the spatial coordinates. This is, in
fact, a general consequence of assuming a scaling form $\sigma_{ij}=z
f_{ij}(r/z)$ for the stresses, valid even when the angle of the
sandpile is below the angle of repose. Since this is not proved in
Ref.~\cite{goddard}, we outline the proof here.

Choosing units in which the gravitational force per unit
volume $\rho g$ is unity, the force balance equations for the pile are
\begin{eqnarray}
\partial_r \sigma_{rr}+\partial_z\sigma_{rz}&=&0\nonumber\\
\partial_r\sigma_{rz}+\partial_z\sigma_{zz}&=&1.
\label{fbal1}
\end{eqnarray}
With a scaling form for the stresses, these yield
\begin{eqnarray}
f_{rr}^\prime+f_{rz}-S f_{rz}^\prime&=&0\nonumber\\
f_{rz}^\prime+f_{zz}-S f_{zz}^\prime&=&1
\label{fbal2}
\end{eqnarray}
where $S=r/z$ is the argument of the scaling functions and the prime
indicates differentiation with respect to $S$.
In an elastic region, the fact that the stresses are related to
underlying strain fields requires the consistency condition
\begin{equation}
\partial_z\partial_z\sigma_{rr}+\partial_r\partial_r\sigma_{zz}
=\partial_r\partial_z\sigma_{rz}
\label{consist1}
\end{equation}
which with the scaling form for the stresses is equivalent to
\begin{equation}
S^2 f_{rr}^{\prime\prime}+f_{zz}^{\prime\prime}=
-S f_{rz}^{\prime\prime}.
\label{consist2}
\end{equation}
Differentiating Eqs.(\ref{fbal2}), together with Eq.(\ref{consist2}),
we have three linear homogenous equations in the three second
derivatives $f_{ij}^{\prime\prime}.$ Since the condition for the three
equations to be linearly dependent is $1+S^2+S^4=0,$ which is not
satisfied for any real $S,$ it follows that $f_{ij}^{\prime\prime}$
must all vanish, {\it i.e.\/} the stresses in any elastic region are
linear functions of the spatial coordinates. Since all the stresses
must vanish on the outer surface of the sandpile, it follows that if
the outermost region were elastic, the stresses would be of the form
$\sigma_{ij}=k_{ij} z (1-S/S_0),$ where $r/z=S_0$ on the outer surface
of the sandpile.  Thus the ratios of the stresses to each other are 
independent of spatial coordinates in this region.
From Eq.(\ref{M-C}), it is then clear that if the
Mohr-Coulomb criterion is satisfied as an inequality (as it must
inside an elastic region), it will be satisfied as an inequality even
on the inner boundary of the region, so that one cannot patch the
solution to a plastic region.  Since it is impossible~\cite{goddard} to
obtain a fully elastic solution for the stresses in a sandpile, the
outermost region must be plastic.

In a plastic region, Eq.(\ref{consist2}) will not be satisfied.
However, Eqs.(\ref{fbal2}) are still valid. For the outermost region,
this implies that the scaling functions $f_{ij}(S)$ are all linear in
$S-S_0$ close to the surface of the pile. The proportionality constants
can be obtained from Eq.(\ref{M-C}) with Eqs.(\ref{fbal2}). But since
Eqs.(\ref{fbal2}) with the derivative of Eq.(\ref{M-C}) constitutes a
system of first order differential equations in the scaling functions
$f_{ij},$ one can evolve them to obtain the stresses everywhere in the
outermost plastic region once they are known at the surface. Since
choosing $f_{ij}$ to be a linear function of $S-S_0$ can be seen to be
a valid solution not just close to the surface of the pile, but even as
one proceeds into its interior, it follows that this is the correct
solution.

Thus we see that for {\it any\/} angle for the sandpile, the outermost
region must be plastic, and the stresses are linear homogenous
functions of $(z-r/S_0).$ The extent of this plastic wedge is a free
parameter in elastoplastic theories, partly constrained by the
requirement of matching to an elastic region. Since the ratios of the
stresses to each other are independent of location in the plastic zone,
it follows that the yield lines, along which the system destabilizes
under infinitesimal extra shear, have the same orientation everywhere
in the outermost plastic zone. For the case when the pile is at the
angle of repose, the yield lines are vertical and along the surface of
the pile. If the plate supporting it at the bottom is made of segments,
and an infinitesimal extra upward force is applied to a segment below
the plastic region of the pile, the segment will move upwards. If the
extra force is kept constant, independent of the displacement of the
segment, the sandpile will not be able to resist the force, and the
segment will continue to move upwards, resulting in overflow at the top
surface. This is in contrast to the response that one would obtain in
the elastic regions, where an infinitesimal extra applied force would
cause a slight upward motion, at which point the (deformed) sandpile
would be able to resist the extra force.  Note that such an argument
relies on the notion that applying the extra force will cause the
material in contact with the segment that is pushed to move upwards,
propagating the disturbance upwards without attenuation.  Implicit in
this argument is the idea that stresses can be related to strains;
since the approach of Refs.~\cite{bcc} and~\cite{cates} does not
consider strains at all, it is not clear that the same argument would
apply there, which is why we obtained a different experimental test (in
the previous section).

Away from the angle of repose, the yield line in the outer plastic
region in elastoplastic models tilts outwards from the vertical, at an
angle that can be calculated in terms of the angle of the pile and the
angle of repose. The above argument would apply there too, with the
obvious modification that the extra applied force would have to be
oriented along the yield direction.

\noindent{\bf 4. Experimental Consequences}

We now examine the effect on experiment of various approximations we have made:
setting the Bagnold hysteresis angle to zero, applying an infinitesimal
force to a perfectly continuous medium, and considering a
two-dimensional sandpile.

Real granular materials do not have a well
defined angle of repose. If one pours grains slowly from a point
source, the pile slowly builds up to a maximum angle; adding any more
grains causes an avalanche in which the pile suddenly collapses to an
angle that is about ten per cent less. If one keeps adding grains
beyond this, the pile builds up once again to its maximum angle. The
Bagnold hysteresis angle is the difference between the two angles
before and after the avalanche. We have treated the angle of repose as
determining the maximum shear stress that any part of the pile, regardless
of its past history,  can ever sustain. One should therefore build up the
pile as close as possible to its maximum angle before putting on the
perturbing force. Since the pile will never be {\it exactly\/} at
its maximum angle, one would always need a finite force to destabilize it.
However, a sufficiently large force applied anywhere in the pile will
always destabilize it, so that identifying a region of incipient failure
becomes difficult.

One can resolve this issue statistically: if the
pile is repeatedly brought close to its maximum angle, and a small
force is applied somewhere in the region of incipient failure, the
probability of destabilizing the pile should be independent of where
the force is applied.  One should be able to detect the position of the
line separating the stable region from the outer, unstable, region by
applying the same magnitude perturbation at different distances from
the surface.  Throughout the outer region, the perturbation should be
{\it equally likely\/} to make the pile fail.  As soon as the position
of the perturbation crosses into the inner region, the perturbation
will become progressively less likely to cause failure as the distance
from the separating line increases. Thus, if only a thin boundary layer
is on the verge of flowing, the probability of destabilizing the pile
will decrease rapidly when the point of application of the force is
moved away from the surface.

The applied force should also be sufficiently small so as not to
destabilize a region well below the Mohr-Coulomb yield point; it is
preferable to distribute the force over several grains, since a highly
localized force will produce much greater stresses in its vicinity.
Finally, if one constructs a two-dimensional pile by pouring grains
between parallel plates, the friction between the plates and the grains
should be sufficiently small so as to avoid shear forces at the boundary
plates stabilizing the pile.

These issues are much less important for elastoplastic theories, where
the extra force is applied across an extended segment of the bottom
plate, and the experiment can be performed even away from the maximum
angle of repose.  Of course, one would need to know the angle of the pile quite
accurately in order to apply the perturbing force in the correct
direction.

In conclusion, we have shown that the existence of a wide failure zone
adjoining the surface of a two-dimensional sandpile built to the angle
of repose is amenable to experimental verification. The presence of such a
zone is perhaps surprising.  That it should be predicted by two very
different theories is remarkable and deserves a careful experimental study.

\noindent{\bf Acknowledgments}
We thank Michael Cates, Joe Goddard, Dov Levine and Tom Witten and for
many stimulating and enlightening discussion about plasticity and
fragility in sandpile models. The authors are grateful to the Institute
for Theoretical Physics at Santa Barbara for their hospitality where
this work was initiated. SRN was supported by the NSF under Award
CTS-9710991 and by the MRSEC Program of the NSF under Award
DMR-9400379. ON was supported by the Alfred P. Sloan Foundation.

\end{multicols}

\begin{references}
\bibitem{caveat1}
Both types of theories (Refs.~\cite{goddard} and~\cite{bcc}) ignore the
Bagnold hysteresis angle. This
means that in those theories a pile can be built up to an angle of repose,
and any
material added subsequently will simply cause the pile to grow
laterally, keeping the sandpile slope fixed at the angle of repose. In
fact, the motion of the grains in the boundary layers lowers
the inclination of the sandpile by the Bagnold hysteresis angle.  Since this
is approximated to zero in both approaches, one would expect the boundary
layer
to be of negligible width to within the same approximation.
The implications for experiment of this assumption will be
considered later.
\bibitem{caveat3}
This can be achieved by choosing the perturbation $f$ to be small, and
rounding it out so that it is not concentrated at a point.
\end{references}
\end{document}